\documentclass[prl,twocolumn,showpacs,amsmath,amssymb]{revtex4}
\usepackage{graphicx}
\usepackage{dcolumn}
\usepackage{bm}
\begin{document}
\newcommand{\cm}{cm$^{-1}$}
\newcommand{\sig}{$\sigma(\omega)$}
\newcommand{\rp}{$R_p$}
\title{Heavy Carriers and Non-Drude Optical Conductivity in MnSi}
\author{F.~P.~Mena$^{1}$, D.~van~der~Marel$^{1}$, M.~F\"{a}th$^{2}$, A.~A.~Menovsky$^{2}$ and J. A. Mydosh$^{2,3}$}
\address{$^{1}$Material Science Center, University of Groningen, 9747 AG Groningen, The Netherlands.
$^{2}$Kamerlingh Onnes Laboratory, Leiden University, 2500 RA
Leiden, The Netherlands. $^{3}$Max-Planck-Institute for Chemical
Physics of Solids, D-01187 Dresden, Germany.}
\date{november 11, 2002}
\begin{abstract}
Optical properties of the weakly helimagnetic metal MnSi have been
determined in the photon energy range from 2 meV to 4.5 eV using
the combination of grazing incidence reflectance at $80^\circ$ (2
meV to 0.8 eV ) and ellipsometry (0.8 to 4.5 eV). As the sample is
cooled below 100 K the effective mass becomes strongly frequency
dependent at low frequencies, while the scattering rate developes
a linear frequency dependence. The complex optical conductivity
can be described by the phenomenological relation
$\sigma(\omega,T)\propto (\Gamma(T)+i\omega)^{-0.54}$ used for
cuprates and ruthenates.
%
%
%
%
\end{abstract}
\maketitle

\narrowtext


%
%
\begin{figure}
  \centerline{\includegraphics[width=7cm,clip=]{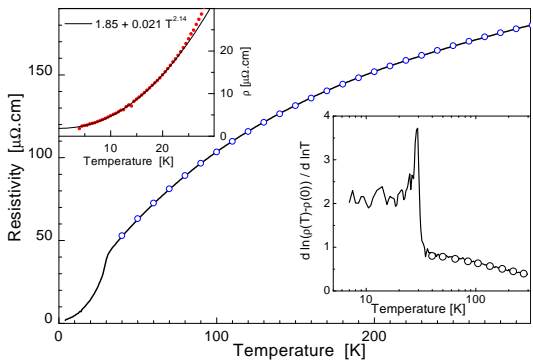}}
  \caption{
  DC resistivity as a function of temperature (solid curve). The open symbols
  represent $\rho_{p}(T)=( 1 / \rho_{\infty} + 1/ (AT) )^{-1}$ with
  $\rho_{\infty} = 286 \mu\Omega cm$ and $A=1.62\mu\Omega cm K^{-1}$.
  Top left inset: DC resistivity below 30 K (dots) and fit to $\rho_F(T)=\rho(0)+AT^{\mu}$.
  Lower right inset: Temperature dependence, $\mu(T)$, of the exponent in $\rho(T)=\rho(0)+AT^{\mu}$ (solid curve).
  The open symbols represent $dln\rho_{p}/dlnT$, where $\rho_{p}(T)$ is the same function as
  in the main panel.
  }\label{resistivity}
\end{figure}
\begin{figure}
  \centerline{\includegraphics[width=7cm,clip=]{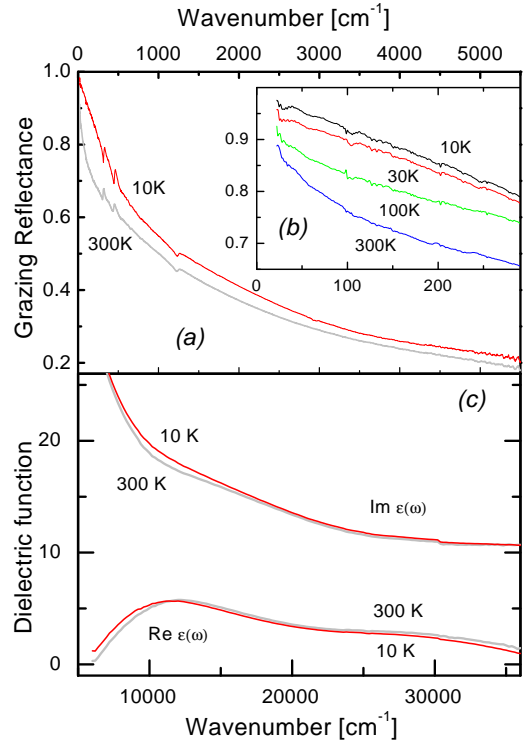}}
  \caption{
  {\it a)} Grazing reflectivity at 80$^\circ$ angle of incidence measured at 10 and 300 K
  {\it b)} Expanded view of the reflectivity below 300 \cm.
  {\it c)} Real and imaginary part of the dielectric function in the visible part of the
  spectrum measured with spectroscopic ellipsometry.
  }\label{reflectivity}
\end{figure}
\begin{figure}
  \centerline{\includegraphics[width=7cm,clip=]{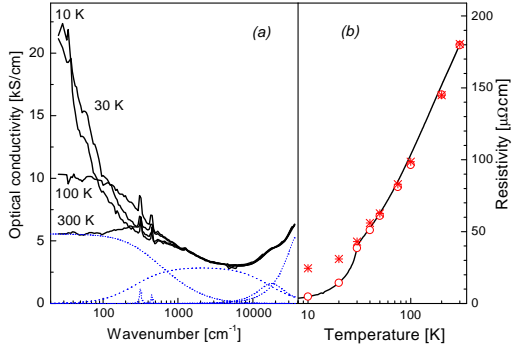}}
  \caption{
  {\it a)} Optical conductivity at four different temperatures (solid lines) and components of the non-Drude plus Lorentz
  oscillators fit at 300 K (dotted lines).
  {\it b)} Measured DC resistivity, and DC resistivity obtained
  by extrapolating the experimental $\sigma(\omega)$ using a
  Drude-Lorentz fit (stars) and using Eq. \ref{NFL-sig} (open circles). The
  fit-parameters are presented in Fig. \ref{param}.
  }\label{sigma}
\end{figure}
\begin{figure}
  \centering
  \includegraphics[width=7cm,clip=]{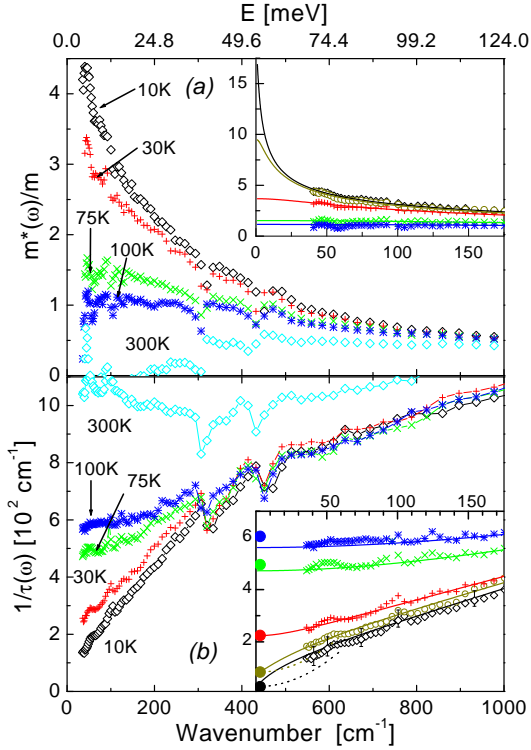}
  \caption{
  {\it a)} Effective mass ({\it Inset:} Behavior below 200 \cm\
  for 100, 75, 30, 20, 10 K starting from below)
   and {\it (b)} frequency-dependent-scattering rate ({\it Inset:}
   Behavior below 200 \cm\ for
  100, 75, 30, 20, 10 K starting from above)
  in MnSi as obtained from \sig\ at different temperatures.
  Symbols represent the experimental data and thick lines the calculation from the
  non-Drude fit described in the text. The solid points at the
  left show $\rho_{DC} \omega_p^2 / (4 \pi)$. The inset of the
  lower panel shows also the expected frequency dependence in the
  Fermi liquid theory calculated from Eq. \ref{tauFL} (dotted lines). This
  dependence is not compatible with our measured $1/\tau(\omega)$
  and cannot be explained with errors coming from the measured
  reflectivity as shown by the error bars.
  }\label{msm}
\end{figure}
\begin{figure}
  \centerline{\includegraphics[width=7cm,clip=]{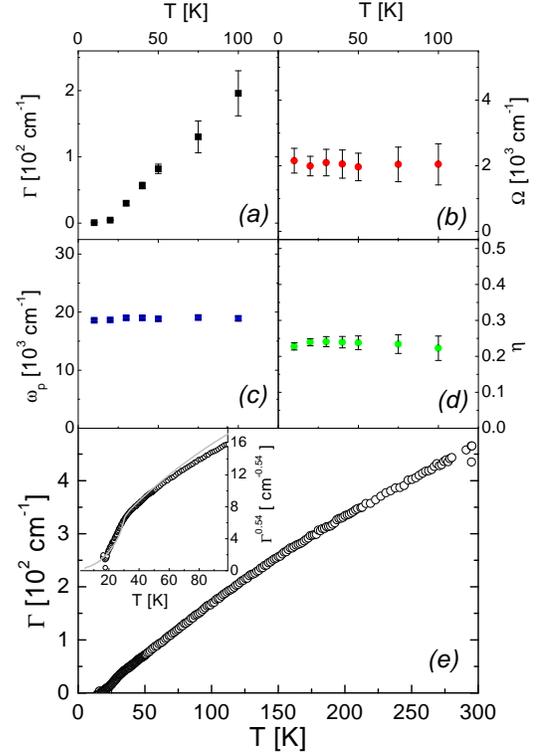}}
  \caption{
  Parameters for the non-Drude optical conductivity, Eq.
  \ref{NFL-sig}, obtained by fitting the complete set of data (reflectivity, ellipsometry and DC resistivity):
  {\it a)} $\Gamma$, {\it b)} $\Omega$, {\it c)} $\omega_p$ and  {\it d)} $\eta$.
  {\it e)} Temperature dependence of $\Gamma$ (see text). {\it Inset:} Temperature dependence of
  $\Gamma^{0.54}$ (circles) and of $\rho_{DC} \omega_p^2/(4 \pi \Omega^{2\eta})$ (solid line).
  }\label{param}
\end{figure}
The weakly helimagnetic metal MnSi ($T_C=29.5$ K) has been the subject of intensive studies during the last 40
years. In the helimagnetic phase the resistivity has a $T^2$ dependence, which has been explained as resulting
from a coupling of the charge carriers to spin fluctuations\cite{moriya}. Recently interest has shifted to the
quantum phase transition \cite{pfleiderer,pfleiderer-nature} at a critical pressure of $14.6$ kbar where the
Curie temperature becomes zero. The temperature dependence of the resistivity outside the magnetically ordered
region, at high pressures, has been found to be proportional to $T^{3/2}$ in temperature range far larger than
that predicted by the so-called nearly ferromagnetic Fermi-liquid theory (an extension of the Fermi-liquid
picture)\cite{pfleiderer-nature}. This fact has suggested the non-Fermi liquid nature of MnSi in the normal
state \cite{pfleiderer-nature}. Despite these efforts in understanding the physics behind MnSi, few attempts
have been made to determine and understand its optical properties. Measurements below $T_C$ of the far infrared
normal incidence reflectivity indicated a remarkable departure from the Hagen-Rubens law, usually observed in
metals\cite{damascelli}. However, the high value of the reflection coefficient close to the 100 $\%$ line
prevented a detailed analysis of the frequency dependent optical conductivity in this range. In this Letter we
overcome this hurdle by using $p$-polarized light at a grazing angle of incidence of $80^{\circ}$, for which the
reflection coefficient drops well below the 100 $\%$ line. We show that the frequency dependent scattering rate,
and the effective mass deviate from the behavior expected for Fermi liquids which can be described with an
expression for \sig\ that departs from the usual Drude model.


Single crystals were grown using the travelling floating zone
technique\cite{damascelli,fath}. The temperature dependence of the
resistivity is shown in Fig. \ref{resistivity}. Fitting the
resistivity to the equation $\rho(T) = \rho(0) + A T^{\mu}$ in the
temperature interval 4K to 23K, we obtain  $\rho(0)=1.85
\mu\Omega$.cm, $A = 0.021 \mu\Omega$.cm.K$^{-\mu}$, and $\mu=2.1$.
The resistivity increases more rapidly in the region between 23 K
and the phase transition. For $T>30$K the resistivity fits to the
formula $\rho_{p}(T)=(1/\rho_{\infty}+1/(A T))^{-1}$ with
$\rho_{\infty}=286\mu\Omega$cm and $A=1.62\mu\Omega$cmK$^{-1}$.
The remarkable accuracy of this description is further confirmed
by the logarithmic derivative shown in the inset of Fig.
\ref{resistivity}. The tendency of the resistivity toward
saturation at a value $\rho_{\infty}$ for $T\rightarrow\infty$ is
in agreement with Calandra and Gunnarsson's
result\cite{gunnarsson} that the resistivity saturates when the
mean free path $l=0.5n^{1/3}d$ (roughly the Ioffe-Regel limit),
where $n$ is the density of the electrons and $d$ is lattice
parameter. Also this indicates that, if the temperature saturation
would be absent, the resistivity would be proportional to
$T^{\mu}$, where the exponent $\mu=1.0$ with a very high accuracy.
These observations stand in stark contrast to the $T^{5/3}$
temperature dependence predicted from the model of
spin-fluctuations in itinerant electron magnetism\cite{moriya}.
Yet the overall temperature dependence, and the strong reduction
of $\rho(T)$ below $T_C$ indicate a dominant electronic (or spin)
contribution to the scattering mechanism.

Grazing incidence reflectivity was measured in the range 20 to
6000 \cm\ using a Bruker 113v FT-IR spectrometer (see Fig.
\ref{reflectivity}a and \ref{reflectivity}b). The temperature
dependence was measured using a home-built cryostat, the special
construction of which guarantees the stable and temperature
independent optical alignment of the sample. The intensities were
calibrated against a gold reference film which was evaporated {\it
in situ} without repositioning or rotating the sample-holder. In
the range 20 to 100 \cm\ we measured the temperature dependence of
the grazing reflectivity with 0.5 K intervals below 50 K and 2 K
intervals above 50 K. The complex dielectric function in the range
6000 to 36000 \cm\ was measured with a commercial (Woollam VASE32)
ellipsometric spectrometer for the same set of temperatures as the
grazing reflectivities using an ultra high vacuum cryostat (shown
for 10 K and 300 K in Fig. \ref{reflectivity}c). The complex
dielectric function $\epsilon(\omega) = \epsilon'(\omega) + i
(4\pi/\omega)\sigma_1(\omega)$ was calculated from the complete
data set (grazing infrared reflectance and visible ellipsometry)
using Kramers-Kronig relations, following the procedure described
in Ref. \onlinecite{eb}. Below 20 \cm\ the reflectivity data were
extrapolated to fit the experimentally measured DC conductivities.
The optical conductivity is shown for some temperatures in Fig.
\ref{sigma}.

The first remarkable feature in the spectra is the similarity of
the optical conductivity to the response of heavy fermion
systems\cite{degiorgi}. In those materials, \sig\ has almost no
temperature dependence down to a frequency of $\sim$10 \cm\ and,
below this frequency, a narrow mode centered at zero frequency is
formed \cite{degiorgi}. Similar behavior has also been noticed for
$\alpha$-cerium\cite{eb} in the mid-infrared frequency range. 
Following a common procedure in the study of the electrodynamic
response of heavy fermion systems, we have calculated
$1/\tau(\omega)$ and $m^\star(\omega)/m$ from the optical
conductivity using the extended Drude-model \cite{hopfield}:
\begin{equation}\label{ext-drude}
  \sigma(\omega)=\frac{\omega_p^2}{4\pi}\frac{1}{1/\tau(\omega)-i\,\omega\, m^\star(\omega)/m}
\end{equation}
adopting the value $\omega_p/2\pi c = 18700$\cm\ for the plasma
frequency, motivated by the least square fits which we will
discuss below. Fig. \ref{msm}a indicates a significant mass
renormalization at low frequencies which, at the lowest measured
temperatures, shows no indication of reaching a frequency
independent value. Previous de Haas-van Alphen experiments (at
$T=$0.35 K) \cite{taillefer} provided an {\em average} enhancement
of 4.5 times the cyclotron mass, although values as high as 14
were observed for some of the orbits. This average value was found
to be compatible with the enhancement of the linear coefficient of
the heat capacity $\gamma/\gamma_o=$5.2 calculated from specific
heat data of Ref. \onlinecite{fawcett}. In comparison, our data
show at 10 K and at the lowest measured frequency an enhancement
of 4, and an enhancement of 17 when we extrapolate the data to
$\omega = 0$.

The second remarkable feature is the behavior of $1/\tau(\omega)$
(Fig. \ref{msm}b). At high temperatures this quantity becomes
frequency independent, as expected for a Drude peak. Already at
100 K $1/\tau(\omega)$ is no longer a constant. Approaching the
phase transition $1/\tau(\omega)$ becomes strongly frequency
dependent between 30 and 300 \cm\, and it follows approximately a
linear frequency dependence in this frequency range. In contrast,
other correlated systems, such as heavy fermions
\cite{degiorgi,eb} and perovskite titanates \cite{katsufuji}, show
a frequency-dependent scattering rate with an $\omega^2$
dependence at low frequencies. Indeed the theory of Fermi liquids
\cite{pines} predicts
\begin{equation}\label{tauFL}
    1/\tau(\omega,T) = 1/\tau_o + a (\hbar\omega)^2 + b (k_B T)^2
\end{equation}
with  $b/a = \pi^2$. The same expression was obtained by Millis
and Lee considering the Anderson lattice model \cite{millis}, and
qualitatively similar behavior has been calculated by Riseborough
in the context of spin-fluctuations\cite{riseborough}. MnSi has
also a $T^2$ dependence of the DC resistivity below $T_C$ and the
corresponding expected frequency dependence of $1/\tau(\omega)$ is
plotted in the inset of the lower panel of Fig. \ref{msm} (dotted
lines) for 10 and 20 K. There is a mismatch with the experimental
$1/\tau(\omega)$, both in absolute value and the observed trend,
which is outside the experimental error bars. However it can not
be excluded, that at an even lower frequency the experimental
$1/\tau(\omega)$ would cross over to a $\omega^2$-dependence.

Above, we have pointed out various striking results in the optical
response of MnSi. In order to understand their nature, let us take
a closer look at the low frequency data. From 300 to 75 K,
$1-R_p(\omega)$ follows a $\omega^{1/2}$ behavior (see Fig.
\ref{reflectivity}b). This can be easily understood from the fact
that at low frequencies, from the Fresnel formulae, \rp\ can be
written approximately as:
\begin{equation}\label{rp}
  R_p=1-\frac{2 \omega^{1/2}}{\cos \theta} {\mbox
  Re}\left[\frac{1}{\sqrt{i\pi\sigma(\omega)}}\right]
\end{equation}
where $\theta$ is the angle of incidence. In the case that
$\sigma_1$ is constant and $\sigma_2$ goes to zero, this
expression reduces to the well known Hagen-Rubens law. In the
Drude picture this corresponds to the frequency range where the
scattering rate is larger than $\omega$. In contrast, below 75 K
our measured $R_p$ does not follow a $\omega^{1/2}$ behavior.
Combining the Drude model with the Fresnel equations for
reflectivity, a plateau in the reflectivity is expected for {\it
intermediate} frequencies (frequencies larger than the scattering
rate but much lower than the plasma frequency). To check this more
closely we measured \rp\ below 100 \cm\ in a finer temperature
mesh. Our results show no sign of a plateau, instead
$1-R_p(\omega)$ evolves gradually to a linear frequency dependence
when T is lowered. We can then conclude that either the peak
centered at zero frequency departs from the Drude picture or other
modes appear at low temperatures and at low frequencies.

To distinguish between these alternatives we have fitted, simultaneously, the measured reflectivity,
ellipsometry and resistivity with two models. First we modelled the data with a Drude peak and a set of
oscillators. In this case, the fit fails to reproduce the measured DC resistivities at low temperatures (stars
in Fig. \ref{sigma}b). On the other hand, if we give more fitting weight to $\rho_{DC}$, the result is a poor
fit of $R_p$ at low frequencies. Although this can in principle by remedied by introducing an arbitrary number
of oscillators at frequencies below 100 \cm, the infrared properties together with the DC resistivity can be
summarized in an economical way ({\em i.e.} requiring a minimal set of adjustable parameters) when we replace
the Drude formula with\cite{marel-nfl}
\begin{equation}
    \sigma(\omega) = \frac{\omega_p^2}{4\pi}
    \frac{i}{(\omega+i\Gamma)^{1-2\eta}(\omega+i\Omega)^{2\eta}}
    \label{NFL-sig}
\end{equation}
which for $\eta=0.25$ corresponds to the model by Ioffe and Millis\cite{ioffe-millis}, and for $\Gamma \ll
\omega \ll \Omega$ to Anderson's\cite{pwa} power-law formula $\sigma(\omega) \propto (i\omega)^{2\eta-1}$, both
in the context of the optical conductivity of the cuprate high T$_c$ superconductors.  Eq. \ref{NFL-sig}, in the
case $\Omega \gg \omega$, has been shown to describe the optical conductivity of SrRuO$_3$, below 40 K, in the
range [6-2400]cm$^{-1}$ with $\eta=0.3$ \cite{dodge}. For SrRuO$_3$ this behavior has been justified as arising
from the coupling of electrons to orbital degrees of freedom\cite{laad}, and in the context of the discrete
filamentary model of charge transport\cite{phillips}.

Our new fit, non-Drude plus Lorentz oscillators (whose individual
components at 300 K are displayed in Fig. \ref{sigma}a), gives the
same overall result at high temperatures ($T>$75 K) as the Drude
fit. However, at low temperatures, the non-Drude equation gives a
better fit at low frequencies and, what is more important,
reproduces $\rho_{DC}$ at all temperatures (open symbols in Fig.
\ref{sigma}b). Therefore, we conclude that the low frequency
optical response of MnSi is best described by Eq. \ref{NFL-sig}.
From the fit we can extrapolate the optical properties to lower
frequencies (insets of Fig. \ref{msm}). The extrapolation shows
that at 10 K, for $\omega\rightarrow 0$, $m^\star(\omega)/m=17$,
with a gradual decay as a function of increasing frequency.
Similarly, $1/\tau(\omega)$ is approximately proportional to
$\omega$ in the frequency range below 300 \cm. Above $T_C$, it has
a weak $\omega^2$ frequency dependence.

Now let us analyze the parameters of the non-Drude conductivity as
provided by the fit to Eq. \ref{NFL-sig}. These values are
summarized in Fig. \ref{param}a-d where the error bars represent
the interval of confidence calculated for a variation of 1\% of
$\chi^2$. Within those error bars, the parameters $\omega_p$,
$\Omega$ and $\eta$ are temperature independent, which contrasts
with the strong temperature dependence of $\Gamma$. This has
another interesting consequence in connection with the DC
resistivity. From Eq. \ref{NFL-sig} we can easily see that
$\rho_{DC} = 4 \pi \omega_p^{-2} \Omega^{2\eta} \Gamma^{1-2\eta}$,
but since $\omega_p$, $\Omega$ and $\eta$ are temperature
independent, $\rho_{DC}(T) \propto \Gamma(T)^{1-2\eta}$. For our
sample, using the values of $\omega_p=18867$ \cm, $\Omega=2049$
\cm\ and $\eta=0.23$ (from Fig. \ref{param}), we obtain
$\rho_{DC}=6.02\, \Gamma^{0.54}$ [$\mu\Omega$.cm]. Recently, Dodge
{\em et al.}\cite{dodge} have emphasized a similar non-linear
relationship between the DC resistivity and the parameter $\Gamma$
in the case of the weak itinerant ferromagnet SrRuO$_3$. The
conclusions for SrRuO$_3$ have been questioned recently by Capogna
{\em et al.}\cite{capogna}, who argued that the true temperature
dependence of the optical properties may have been masked by the
large residual resistivity of the sample used in Ref.
\onlinecite{dodge}. In the present work this problem is absent due
to the low residual resistivity of single crystalline MnSi. In
fact, we can go a step further and try to give a detailed picture
of the temperature dependence of $\Gamma$. For that purpose we fit
the low frequency \rp\ (at all the measured temperatures) to Eq.
\ref{NFL-sig} using the known values of $\omega_p$, $\Omega$ and
$\eta$. The values obtained for $\Gamma$ are displayed in Fig.
\ref{param}e. The inset shows $\Gamma^{0.54}$ and
$\rho_{DC}/6.02$. We can see that the model represented by Eq.
\ref{NFL-sig} describes the measured data (reflectivity and
resistivity) down to approximately 20 K. Below this temperature a
fit only to reflectivity produces unphysical negative values for
$\Gamma$. However, introducing the measured $\rho_{DC}$ produces a
$\chi^2$ which is not more than twice that obtained when fitting
only reflectivity. Apparently at low temperatures there are still
details which Eq. \ref{NFL-sig} is not able to describe.

At low frequencies, deviations from the Drude formula of the
optical conductivity have been seen accompanied by deviations from
$T^2$ in $\rho_{DC}$. Well known examples are YBCO
\cite{schlesinger} and more recently CaRuO$_3$ \cite{lee}.
Therefore, a departure from Drude behavior has been usually
considered as evidence against Fermi-liquid behavior. Here, for
MnSi, we are confronted with an atypical case. The resistivity has
a quadratic temperature dependence, but the optical conductivity
is better described by Eq. \ref{NFL-sig} with $\eta\approx 0.23$,
a clear departure from the Drude formulation. Moreover, instead of
an $\omega^2$-type frequency dependent scattering rate, which is
usually observed in strongly interacting Fermi-liquids
\cite{eb,degiorgi}, here $1/\tau(\omega)$ has a linear frequency
dependence. Although Eq. \ref{NFL-sig} summarizes in a compact way
the low frequency optical response, differing in a fundamental way
from conventional Drude behavior, its microscopic origin is as yet
not fully understood.

For frequencies below 300 \cm\ and for $T < 100$ K the situation
can be summarized as follows: (i) $m^*/m$ decreases from ~17 to ~1
as temperature and frequency are increased. (ii)
Phenomenologically the DC conductivity and the optical
conductivity follow $\sigma \propto (\Gamma(T)+i\omega)^{-0.5}$.
In this formulation $\Gamma(T)\propto T^4$ below $T_C$, whereas
above T$_c$ the temperature temperature dependence is
approximately linear. (iii) For $T>T_C$ the scattering rate
$1/\tau(\omega,T)$ is proportional to $T$ and $\omega^2$ in
contradiction with the theory of weak itinerant
ferromagnetism.(iv) For $T<T_C$ the scattering rate is
proportional to $T^2$ and $\omega$. Given the frequency range for
this type of measurements, we can not exclude the possibility,
that for frequencies below 30\cm\ the scattering rate crosses over
to the Fermi-liquid result $1/\tau\propto\pi^2T^2+\omega^2$.

This project is supported by the Netherlands Foundation for
Fundamental Research on Matter with financial aid from the
Nederlandse Organisatie voor Wetenschappelijk Onderzoek.

\end{document}